\documentclass[a4paper,letters,usenatbib]{mnras}

\usepackage{mathptmx}
\usepackage{txfonts}

\usepackage[T1]{fontenc}
\usepackage{ae,aecompl}

\usepackage{graphicx}	% Including figure files
\usepackage{amssymb}	% Extra maths symbols
\usepackage{rotating}
\title[A discovery of YSOs in older clusters of the LMC]{A discovery of young stellar objects in older clusters of the Large Magellanic Cloud}

\author[B.-Q. For and K. Bekki]{
Bi-Qing For,$^{1}$\thanks{E-mail: biqing.for@uwa.edu.au}
and Kenji Bekki$^{1}$
\\
$^{1}$International Centre for Radio Astronomy Research/The University of Western Australia, \\
M468, 35 Stirling Highway, Crawley, WA 6009, Australia}

\date{Accepted XXX. Received YYY; in original form ZZZ}

\pubyear{2016}

\begin{document}
\label{firstpage}
\pagerange{\pageref{firstpage}--\pageref{lastpage}}
\maketitle

\begin{abstract}
Recent studies have shown that an extended main-sequence turn-off is a common 
feature among intermediate-age clusters (1--3 Gyr) in the Magellanic Clouds. 
Multiple-generation star formation and stellar rotation or interacting binaries 
have been proposed to explain the feature. 
However, it remains controversial in the field of stellar populations. Here 
we present the main results of an ongoing star formation 
among older star clusters in the Large Magellanic Cloud. Cross-matching 
the positions of star clusters and young stellar objects has yielded 
15 matches with 7 located in the cluster center. 
We demonstrate that this is not by chance by 
estimating local number densities of young stellar objects for each star cluster. 
This method is not based on isochrone fitting, 
which leads to some uncertainties in age estimation and methods of 
background subtraction. 
We also find no direct correlation between atomic hydrogen 
and the clusters. This suggests that gas accretion for fueling 
the star formation must be happening in situ. 
These findings support for the multiple-generations 
scenario as a plausible explanation for the extended main-sequence 
turn-off. 

\end{abstract}

\begin{keywords}
galaxies: star clusters: general --- Magellanic Clouds
\end{keywords}

%%%%%%%%%%%%%%%%%%%%%%%%%%%%%%%%%%%%%%%%%%%%%%%%%%

%%%%%%%%%%%%%%%%% BODY OF PAPER %%%%%%%%%%%%%%%%%%

\section{Introduction}

Star clusters (SCs) are fundamental building blocks of galaxies. 
Their physical properties encode valuable information not only 
on their own formation processes but also on the evolutionary 
histories of their host galaxies. 
They were once thought to form in a single epoch of star formation that produced 
thousands or millions of coeval stars with the same chemical composition. 
Recent high-quality Hubble Space Telescope (HST) 
photometric studies of color-magnitude diagrams (CMDs)
of intermediate-age (1--3 Gyr) SCs in the Magellanic Clouds 
uncovered evidence of multiple stellar populations with unexpected characteristics, 
such as two distinct main-sequence turn-off for NGC 1846 \citep{Mackey07}. 
Some of the Galactic globular clusters also reveal multiple stellar populations 
with different helium and light element abundances (e.g., see \citealp{Piotto15, Milone16}). 
NGC 2808 is the most extreme cases where this phenomenon was first discovered \citep{Piotto07}. 
These important discoveries  
have challenged the traditional view of SCs being formed in 
a single star formation episode \citep{Mackey08, G15, Bekki09, Milone15}. 

The direct explanation for the 
extended main-sequence turn-off (eMSTO) is that these clusters experienced 
prolonged star formation for $\sim$100-500 Myr \citep{G09, G14}, 
though one Large Magellanic Cloud (LMC) cluster (NGC 1783) has recently been 
reported to have a possible age spread of $\sim$1 Gyr \citep{Li16}.  
These results suggest that LMC clusters that are $\lesssim$1 Gyr 
old might still show signs of ongoing star formation today. 
However, such age spreads have recently been disputed by other observational 
and theoretical studies \citep{Li14, Bastian13}. 
Stellar rotation or interacting binaries that can mimic an age apread 
on the CMD have been proposed as an alternative explanation (\citealp{BD09, Yang13, D15} and references therein). 
It therefore remains unclear if the clusters contain multiple generations of stars. 

To test the prolonged star formation scenario, we adopt a method that does not rely on 
isochrone fitting to the CMDs of the clusters, 
for which results have been refuted due to field star contamination 
in some cases (see e.g., \citealp{CZ16}). 
Young stellar objects (YSOs) are stars in their very early stage and have 
ages of less than 1 Myr \citep{Dunham15} so that the 
detection of YSOs can be considered as evidence of ongoing star formation. 
We therefore search for YSOs in the clusters with ages in the range of 
0.1 to 1 Gyr in the LMC to determine if SCs could host multiple generations of stars.   

In this letter, we report the finding of YSOs 
in the center regions of several older SCs. 
Our finding provides clear evidence for ongoing star formation in SCs that 
once completed their star formation, and therefore demonstrates the presence 
of multiple generations of stars in at least some LMC clusters. 

\section{Analysis and Results}

We employ the HERITAGE band-matched \citep{Seale14} and 
star cluster \citep{Glatt10} catalogues (hereafter S14 and G10) for the analysis. 
The S14 catalogue consists of astronomical objects 
that are detected in multiple HERITAGE images. 
These objects were positionally cross-matched and then further matched to 
\textit{Spitzer} IRAC and MIPS point sources catalogues. 
The S14 catalogue is dominated by bright YSOs 
but potentially contaminated by background galaxies at low flux level. 
The S14 have employed step-by-step cuts to eliminate contamination. 
They also used $Herschel$ photometry constraints on dust mass 
for asymptotic giant branch (AGB) progenitors to distinguish ambigious 
classification between post-AGB and YSOs. 
To take into account the flux confusion limit, YSOs 
were classified as having either a high probable or a moderate possible
likelihood of being a YSO. The S14 catalogue 
employed here contains at most 1$\%$ evolved stars. 

The G10 catalogue lists the derived ages and V-band luminosities of young SCs 
($\sim$9 Myr to 1 Gyr) in the Magellanic Clouds. The G10 study selected SCs from 
a general catalogue of extended objects in the Magellanic System (\citealp{Bica08}; hereafter B08)
and employed the photometric data of the Magellanic Clouds 
Photometric Survey (MCPS; \citealp{Zaritsky02,Zaritsky04}). 
Double or multiple clusters classification \citep{Dieball02} 
is also adopted in the G10 catalogue.

We positionally cross-match the 2493 probable and 1025 possible YSOs listed in the S14 catalogue 
with the 738 SCs of ages in the range of 0.1 to 1 Gyr 
in the G10 catalogue. We adopt a search radius of 10 pc, 
which is a typical cluster size \citep{Piatti14}. 
This search radius corresponds to an angular size of 41.4\arcsec\ by assuming the distance of 
the LMC to be 50 kpc. This radius takes into account the 
uncertainties discussed in the following sections. 

\begin{figure*}
\includegraphics[scale=0.4]{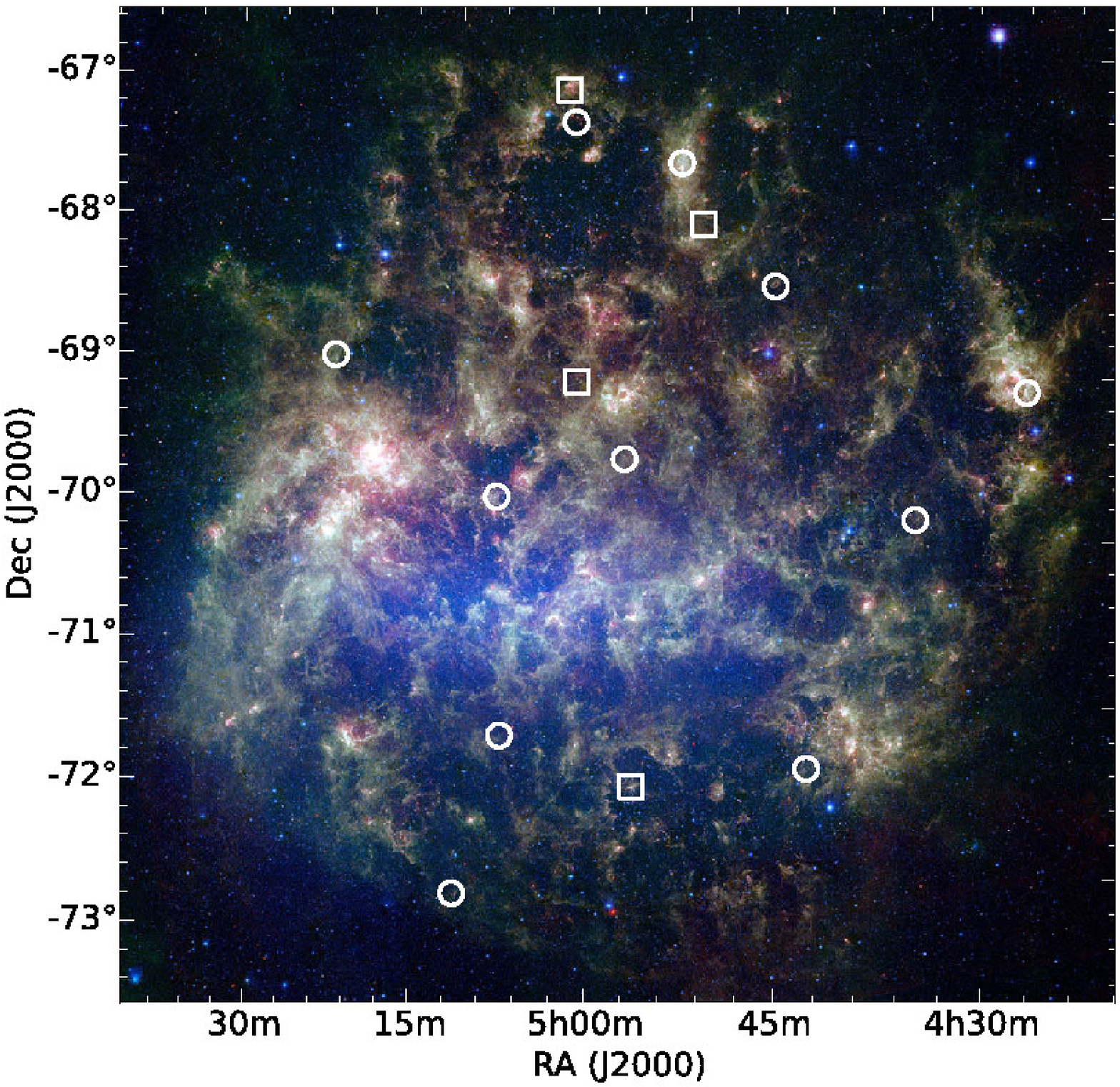}
\includegraphics[scale=0.4]{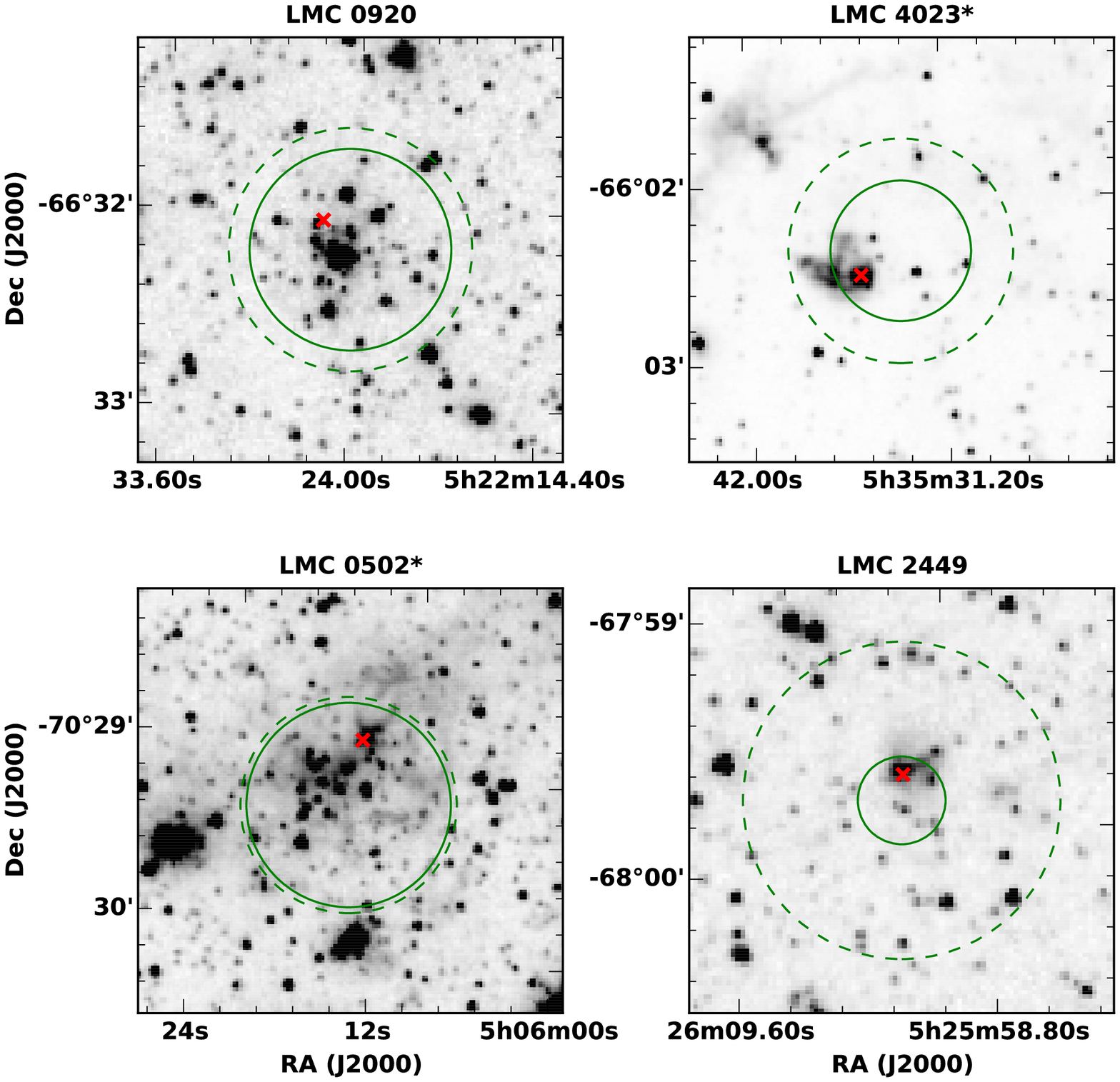}
\caption{Detections of YSOs in the LMC clusters. 
$Left$: Circles and squares show the locations of 15 
star clusters with detections of YSOs in the LMC. 
The clusters are overplotted onto a composite false-color image of 
the LMC from the Spitzer Surveying Agents of Galaxy Evolution survey (SAGE; \citealp{Meixner06}). 
Squares represent the four star clusters shown in the right panel. 
$Right$: Examples of star clusters with detections of YSOs. 
The cluster images are 3.6$\mu$m cutout images from the SAGE survey. 
The solid and dashed circles represent the apparent radii of the 
star cluster and the search radius of 10 pc (41.4\arcsec), respectively. 
Crosses mark the locations of the YSOs. 
An asterisk in the cluster name indicates that the star cluster is a double/multiple system.\label{fig1}}
\end{figure*}

We find 15 probable and 6 possible 
YSOs falling within the search radius. Two of these could also be 
post-AGB stars based on the classification in the S14 catalogue. 
In this Letter, we focus on the 15 probable YSOs, although the possible YSOs could 
also be genuine. Figure~\ref{fig1} shows the locations of the YSO detections 
in the LMC and Table~\ref{tab1} summarizes the cross-matching results and physical parameters 
of the SCs. Five out of fifteen young stellar candidates are confirmed to 
be genuine by photometric and spectroscopic observations, 
such as detection of ice and water masers \citep{GC09, Sewilo10}. 
The other 10 are highly likely to be genuine YSOs being consistent with 
the spectral energy distributions and dust properties of YSOs \citep{Seale14}. 
Among these 15 YSO candidates, 7 are located at the cluster center. 

\begin{table*}
\centering
\begin{minipage}{170mm}
\caption{Summary of cross-matching results and physical parameters of 
the star clusters with detections of young stellar objects\label{tab1}}
\begin{tabular}{lccccccccc} % four columns, alignment for each
\hline
Cluster ID$^{a}$ & 
Cluster RA & 
Cluster Decl. &
$R_{\rm app}$ &
log(age) &
$V^{b}$ &
YSO R.A. &
YSO Decl. &
YSO classification$^{c}$ &
$N^{d}$\\
 & 
deg [J2000] & 
deg [J2000] & 
arcmin & 
yr & 
mag & 
deg [J2000] & 
deg [J2000] & \\
\hline
LMC 0286	&	74.3583	&	$-$67.6853	&	0.50	&	8.60	&	17.65	&	74.356865	&	$-$67.675769	&	defYSO/postAGB	& 3\\
LMC 0502*	&	76.5875	&	$-$70.4794	&	0.65	&	8.20	&	12.87	&	76.593826	&	$-$70.472698	&	probYSO	&         2\\
LMC 0848	&	79.9833	&	$-$68.3500	&	0.16	&	8.00	&	15.71	&	79.960629	&	$-$68.350753	&	probYSO	&         4\\
LMC 0920	&	80.6208	&	$-$66.5283	&	0.57	&	8.60	&	13.68	&	80.629594	&	$-$66.526904	&	probYSO	&         4\\
LMC 2228*	&	79.0250	&	$-$71.8069	&	0.16	&	8.00	&	16.64	&	79.014055	&	$-$71.805417	&	probYSO	&         3\\
\hline
\end{tabular}
Notes. Table 1 is published in its entirely as Supporting Information 
with the electronic version of the paper. A portion is shown here 
for guidance regarding its form and content. (a) Asterisk next to cluster name indicates possible double/multiple populations\citep{Dieball02}. 
(b) $V$ magnitudes correspond to the total luminosity of all stars within $R_{\rm app}$ \citep{Glatt10}. 
(c) post-AGB: post Asymptotic Giant Branch star; defYSO: definite young stellar object; 
probYSO: high probability of being a young stellar object \citep{Seale14}.
(d) Number of YSOs within 100~pc of each star cluster.
\end{minipage}
\end{table*}

\subsection{Reliability}\label{rel}

The reliability of the cross-matching depends on the matching radius, 
completeness of the catalogues, source density and positional accuracy. 
The angular resolution and signal-to-noise ratio of the data sets 
affect the positional accuracy. The S14 catalogue adopts the coordinates of 
point sources in the $Herschel$ PACS 100$\mu$m point source catalogue if available. 
$Herschel$ PACS 100$\mu$m images provide the highest angular resolution of all $Herschel$ wavebands. 
If no detection is documented in the PACS 100$\mu$m point source catalogue, 
the coordinates of the next coarser angular resolution point source catalogue 
are adopted and so on. We examine the positional errors adopted in the S14 catalogue. 
The standard deviation of the errors is 0.1\arcsec\ with a median error of 0.06\arcsec, 
which is very small as compared to the search radius. 

Star cluster coordinates adopted from the B08 have an error of 10\arcsec--15\arcsec, 
which corresponds to a maximum error of 4 pc. The apparent cluster size is generally 
underestimated due to visual inspection of the photographic plates. 
The search radius takes into account 
the positional accuracy of the SC and the 
uncertainty of the apparent cluster size. 
Nevertheless, we verify the matched objects by visual inspection using 
the \textit{Spitzer} IRAC 3.6$\mu$m and $Herschel$ PACS 100$\mu$m images for SCs and YSOs, 
respectively. 

\subsection{Completeness}\label{complete}

There is a selection bias in the G10 catalogue. The catalogue 
only consists of well-defined and not too extended SCs. 
It is also limited to SCs with ages between $\sim$10 Myr and 1 Gyr. 
Star clusters younger than 10 Myr are generally classified as associations or nebulae. 
Thus, they are excluded from the G10 catalogue. The upper age limit is due to the 
limiting photometric magnitude of the MCPS, 
resulting in the difficulty of resolving MSTO points of intermediate-age and older clusters. The S14 catalogue is complete for sources with fluxes brighter than $\sim200$ mJy 
in any single $Herschel$ waveband over $>$ 99$\%$ of the HERITAGE image area with a 
surface brightness of $<$ 100 MJy sr$^{-1}$ in the SPIRES 250$\mu$m waveband. 

\subsection{Probability}\label{prob}

Fifteen SCs ($N_{\rm S-Y}=15$) have YSOs within 10 pc 
of their centres. The total number of SCs ($N_{\rm SC}=738$) 
in the G10 catalogue might be smaller than the actual number 
because of the limited sensitivity of 
the B08 and MCPS. Also, the total number of YSOs ($N_{\rm YSO}=2493$) 
in the S14 catalogue might be significantly smaller than that of all YSOs, 
because only bright YSOs have been detected. Therefore, $N_{\rm S-Y}$ 
is likely to be underestimated. 
However, it is important for the present study 
to investigate whether or not the above $N_{\rm S-Y}$ 
can be explained simply by a high probability of both a YSO and a 
SC being along the same line-of-sight (i.e., chance of alignment) 
and/or due to high local density of stars.
We analytically estimate (i) the local probability for detecting the YSOs within 10~pc of SCs (i.e., search radius) based on 
the number of YSOs within the adopted 100~pc ($N_{\rm YSO-100pc}$) of 15 SCs, and (ii) $N_{\rm YSO-100pc}$ of 15 SCs 
with respect to field star density.

The probability of a YSO being detected within the defined radius ($R$) 
of a SC in the local YSO density ($P_{\rm S-Y}$) 
is $N_{\rm YSO-100pc}\times$($R$ / 100~pc)$^{2}$. The mean probability for $R$ = 10, 3 and 1~pc are 
$5.6\times10^{-2}$, $5.04\times10^{-3}$, and $5.6\times10^{-4}$, respectively, for 15 SCs with YSOs. If we consider the inclination of the LMC ($i$ = 27\degr--45\degr; \citealp{Vdb00}), 
then the projected surface area of the LMC is appreciably smaller but it does not affect $P_{\rm S-Y}$. 
We note that 3 SCs (LMC 2442, LMC 2519* and LMC 3829) have 
a large $N_{\rm YSO-100pc}$ ($>$10). Examining the spatial density of YSOs at 200~pc away from these 3 clusters, we only find a large $N_{\rm YSO-100pc}$ for LMC 2519* and LMC 3829. This suggests that these two SCs are possibly in a very active star forming region and the probability for YSOs that are not physically associated to them is higher. Nevertheless, the low value of 
the mean probabilities demonstrates that detection of the other 13 SCs with YSOs is unlikely to be a chance coincidence 
but must be a real physical association of YSOs with the SCs.

\begin{figure}
\includegraphics[scale=0.3, angle=-90]{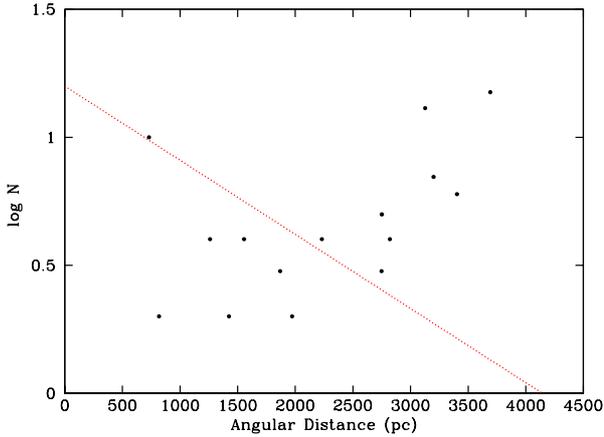}
\caption{Angular distance between the SCs and the LMC centre vs. number of YSOs 
within 100~pc of 15 SCs on logarithmic scale. The dashed line corresponds 
to theoretical prediction of YSO number with respect to angular distance from the LMC centre with an arbitary zero-point for simplication. \label{ndist}}
\end{figure}

By comparing the spatial density of YSOs in each of the 15 SCs with the density profile of the region, 
we can rule out the possibility that the detection of YSOs in these 15 clusters is due to high local field star density. This is because a 
higher detection rate would be expected in the inner region of the LMC, which is not the case. 
The density profile, $\Sigma D$, can be described as exp($-D/l$), where $D$ is the distance between the SCs and the LMC centre 
and $l$ is the scale length of the LMC disc (1.5~kpc).  
In Figure~\ref{ndist}, we show the theoretical density profile (dashed line) and the number of YSOs within 100~pc of SCs as 
a function of distance between the SCs and the LMC centre. There is no correlation.

\section{Discussion and Conclusions}

\begin{figure}
\includegraphics[scale=0.4]{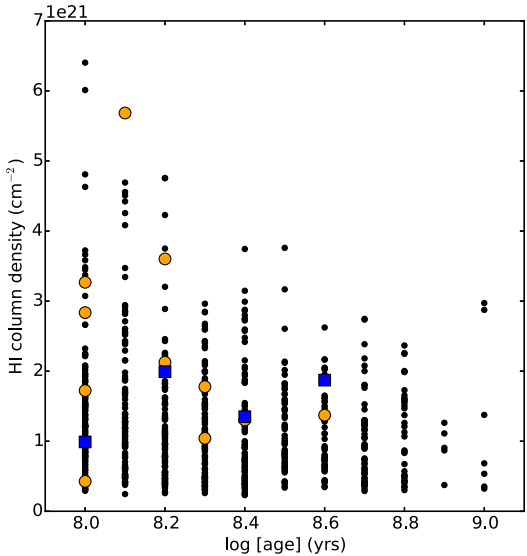}
\caption{Mean HI column density within 10 pc of 738 star clusters vs. cluster age on a logarithmic scale. 
Star clusters with ages between 0.1 to 1 Gyr are selected from the catalogue of young star clusters. 
Circles and squares correspond to the same symbols as shown in Figure 1. 
It is evident that the HI column density decreases with increasing cluster age. 
There is no direct correlation between HI gas and the star clusters with detection of YSOs.\label{higas}}
\end{figure}

\begin{figure}
\includegraphics[scale=0.4]{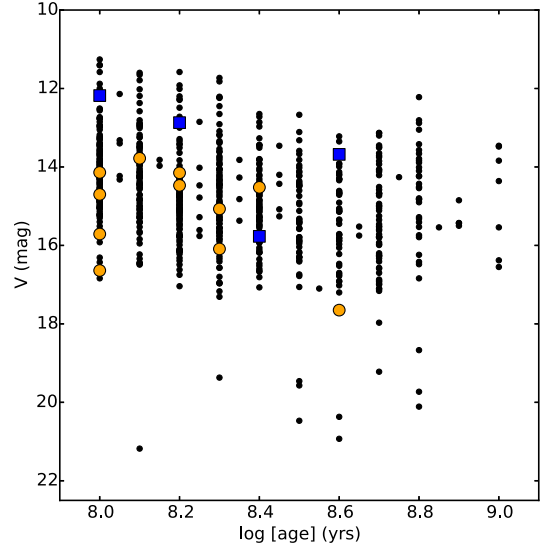}
\caption{Luminosity ($V$ mag) vs. age on a logarithmic scale of star clusters 
with ages between 0.1 to 1 Gyr. Data points and symbols represent the same as 
in Figure 1. There is no detection of YSOs 
in clusters older than 0.4 Gyr old.\label{age}}
\end{figure}

Accretion of interstellar atomic and/or molecular gas onto existing SCs 
can trigger second-generation star formation in the clusters \citep{Bekki09, Li16, PA09}. 
We look for evidence of an external accretion event by investigating the atomic hydrogen gas 
(HI) content in the region of the SCs with identified YSOs using 
the combined single-dish (Parkes) and interferometer (Australia Telescope Compact Array) LMC HI image \citep{Kim03}.
As a SC passes through a region of dense gas, accretion can occur 
and trigger star formation in situ. As shown in Figure~\ref{higas}, 
we find that there is no direct correlation between the HI gas and the clusters, 
which suggests that accretion of interstellar gas onto clusters is 
not responsible for the formation of YSOs in these clusters. 
However, we note that four SCs are situated at the rim of two HI supershells 
(SGS 7 and 11; \citealp{Kim99}). 
These supershells are known to have a relatively high molecular gas concentration \citep{Yamaguchi01}. 
Star formation can be triggered by shell expansion or accretion of the cold molecular gas. 
Interestingly, the detected YSOs in these four clusters reside in the cluster centre. 

While our sample is small, we find that all SCs with YSOs are $\lesssim$0.4 Gyr old (Figure~\ref{age}). 
The lack of YSOs in clusters older than 0.4 Gyr implies that the age difference 
between multiple-generations of stars in these clusters is less than 0.4 Gyr. 
This is consistent with the maximum age differences of multiple stellar populations derived for clusters 
in the LMC \citep{Milone09}. While $\lesssim$0.4 Gyr age spread is commonly 
found among intermediate-age clusters \citep{Mackey08, G15, Milone09}, 
some clusters with similar age to those in our study are consistent with a 
single stellar population \citep{Milone13, D15} or with relatively small age spread 
(i.e., 0.08 and 0.15 Gyr; \citealp{Correnti15, Milone15}). 
Given the accuracy of age estimation for SCs, these are still consistent with 
the present results that the age difference can be $\sim$0.1 Gyr for some clusters. 
Since accretion of interstellar gas onto clusters does not depend on the ages of clusters, 
the result strongly suggests that the required fresh gas supply for the second-generation star 
formation most likely originates from gas ejected by stars inside the clusters. 
All massive stars (more massive than 8 M$_{\odot}$) would have already exploded as supernovae 
in the SCs with ages greater than 0.1 Gyr. The gas ejected from supernovae 
is highly unlikely to still remain within the clusters. 
Some intermediate-mass stars in the clusters are currently dying and ejecting gas through their stellar winds. 
If the ejected gas can be trapped by the gravitational potential of the cluster, star formation from that gas maybe possible.
 
The current star formation rate of the LMC estimated from 299 YSO candidates is 0.06 M$_{\odot}$ yr$^{-1}$ \citep{Whitney08}. 
This rate is a lower limit because of possibly missing YSO candidates in the observation. 
If the star formation rate is scaled to 15 YSOs found in the clusters of our study, 
it corresponds to 0.003 M$_{\odot}$ yr$^{-1}$. This implies that the total mass of stars formed in SCs 
over the last 0.1--1 Gyr can be as large as 2.7$\times$10$^{6}$ M$_{\odot}$. 
This mass is quite significant as compared to the possible total mass of 9.9$\times$10$^{6}$ M$_{\odot}$ 
for 0.1--1 Gyr old SCs with an assumed ($V$-band) mass-to-light ratio of 2.8 in the LMC. 
The large fraction of later generations of stars formed in clusters is comparable 
to the observed fractions of second-generation stars inferred from the 
colour-magnitude diagrams for some LMC clusters \citep{Milone15}. 
Therefore, the second generation of stars in these clusters could have been formed in the 
centre regions of their original clusters. 

Given that dusty YSOs evolve into optically visible pre-main sequence stars (PMS; \citealp{DeMarchi13}), 
the present detection of YSOs implies that there could be PMS stars in the older clusters. 
Pre-main sequence stars have been discovered for several SCs 
in the LMC and Small Magellanic Cloud (SMC) using deep HST 
observations \citep{DeMarchi13, Gouliermis12}, 
though the clusters are younger than those investigated in this paper. 
These PMS stars are very faint ($V > 22$ mag) 
and can not be investigated in existing G10 catalogue derived from ground-based observations. 
If PMS stars were discovered in the old clusters with YSOs, 
then such findings would present an additional irrefutable evidence for the presence of 
multiple-generation of stars in clusters. Since one YSO is detected per SC, 
the star formation rate is 2.0$\times$10$^{-4}$ M$_{\odot}$ yr$^{-1}$ in each cluster. 
This low star formation rate implies that the number of PMS stars in each cluster 
is much lower than those detected in young clusters with initial star bursts. 
Nevertheless, it is worthwhile for future observational studies to investigate 
the numbers and masses of PMS stars in older clusters. 
Such observations would better constrain the ongoing star formation rates in older clusters. 

%A chance event might be the key factor that determines whether a SC can 
%have secondary star formation. The LMC has been strongly interacting with the 
%SMC and our Galaxy over the last 3 Gyr \citep{BC05}, 
%which dramatically changes the physical conditions of the interstellar 
%medium in comparsion with isolated galaxies. This unique situation of the LMC 
%can possibly cause diversity in surrounding gaseous conditions of clusters so 
%that not all clusters experience secondary star formation. 
Our finding also suggests that the gas supply for second-generation star 
formation cannot originate from young massive stars but must be from 
old AGB stars. The presence of low-luminosity clusters ($V\sim17$ mag) in our sample which contains 
YSOs does not provide support for theoretical predictions of a 
threshold mass of globular clusters for second-generation star formation \citep{E08, Bekki11}. 
It is unclear how low-mass clusters can retain ejecta from AGB stars for further star formation. 
However, if low-mass clusters interact with the cold gas from molecular clouds, 
accretion can occur \citep{PA09}. 
Such cold gas accretion might help low-mass SCs to retain some fraction of 
AGB ejecta, though this process needs to be investigated. 

%Our results lend a strong support for the multiple-generation scenario as 
%a plausible explanation for the eMSTO as seen in the LMC clusters. 
%However, the CMDs of some young LMC clusters can be better explained by the sin%gle-generation scenario than the multiple one 
%with stellar rotation as an alternative explanation \citep{Bastian16, PB16}. 
%Therefore, it is too early to conclude that all SCs with eMSTO are due largely to age spreads 
%among cluster stars. Furthermore, the low star formation rate for 
%second-generation stars revealed by the present study strongly suggests that 
%secondary star formation proceeds slowly over a longer time scale, 
%which is consistent with prolonged star formation revealed in intermediate-age LMC clusters 
%\citep{Milone13, Piatti16} but inconsistent with a secondary burst in one cluster \citep{Li16}. 

\section*{Acknowledgements}

This publication makes use of NASA/IPAC archival data 
and data products from the Parkes telescope and the Australia Telescope Compact Array. 
The Australia Telescope Compact Array/Parkes radio telescope 
is part of the Australia Telescope National Facility, which is funded by the 
Commonwealth of Australia for operation as a National Facility managed by CSIRO. 
We thank the referee for his/her constructive comments.  

%%%%%%%%%%%%%%%%%%%%%%%%%%%%%%%%%%%%%%%%%%%%%%%%%%



\bibliographystyle{mnras}
\bibliography{references}

%\begin{thebibliography}{99}
%\end{thebibliography}

% Don't change these lines
\bsp	% typesetting comment
\label{lastpage}
\end{document}